\documentclass[%
reprint,
superscriptaddress,
 amsmath,amssymb,
 aps,
prb,
floatfix,
]{revtex4-1}

\usepackage{comment}
\usepackage{physics}
\usepackage[normalem]{ulem}
\ifdefined\DIRACREP
\else
\newcommand{\DIRACREP}{}

\usepackage{physics}
\usepackage{xparse}
\ifdefined\COSMOMATHS
\else
\newcommand{\COSMOMATHS}{}

\newcommand{\mbf}[1]{\ensuremath{\mathbf{#1}}}

\fi %

\NewDocumentCommand{\rep}{s d<| d|>}{%
\IfBooleanTF{#1}{
   \IfValueTF{#2}{
       \IfValueTF{#3}{\braket{#2}{#3}}{\bra{#2}}
       }{
       \IfValueTF{#3}{\ket{#3}}{}
       }
   }{
   \IfValueTF{#2}{
       \IfValueTF{#3}{\braket*{#2}{#3}}{\bra*{#2}}
       }{
       \IfValueTF{#3}{\ket*{#3}}{}
       }
   }
}

\NewDocumentCommand{\rbra}{sm}{\IfBooleanTF{#1}{\rep*<#2|}{\rep<#2|}}
\NewDocumentCommand{\rket}{sm}{\IfBooleanTF{#1}{\rep*|#2>}{\rep|#2>}}
\NewDocumentCommand{\rbraket}{smom}{
    \IfBooleanTF{#1}{
        \IfNoValueTF{#3}{\rep*<#2||#4>}{\rep*<#2|#3\rep*|#4>}
    }{
        \IfNoValueTF{#3}{\rep<#2||#4>}{\rep<#2|#3\rep|#4>}
    }
}

\NewDocumentCommand{\field}{o m e{_} e{^} o e{_} e{^}}{
\IfValueTF{#5}{\overline{
  #2\IfValueT{#3}{_#3}\IfValueT{#4}{^{\otimes #4}} %
  \otimes
  #5\IfValueT{#6}{_#6}\IfValueT{#7}{^{\otimes #7}} %
  \IfValueT{#1}{;#1}
}}{
  \IfValueTF{#4}{\overline{
     #2\IfValueT{#3}{_#3}\IfValueT{#4}{^{\otimes #4}}
     \IfValueT{#1}{;#1}
  }}
  {#2\IfValueT{#3}{_#3}}
}
}

\NewDocumentCommand{\frho}{o e{_} e{^}}{
\field[#1]{\rho}_{#2}^{#3}
}

\newcommand{\bx}{\mbf{x}}

\newcommand{\e}{a}  %

\NewDocumentCommand{\ex}{e_}{
\IfValueTF{#1}{\e_{#1}\bx_{#1}}{\e\bx}
}  %

\NewDocumentCommand{\lm}{e_}{
\IfValueTF{#1}{l_{#1}m_{#1}}{lm}
}
\NewDocumentCommand{\nlm}{e_}{
\IfValueTF{#1}{n_{#1}\lm_{#1}}{n\lm}
}
\NewDocumentCommand{\enlm}{e_}{
\IfValueTF{#1}{\e_{#1}\nlm_{#1}}{\e\nlm}
}
\NewDocumentCommand{\en}{e_}{
\IfValueTF{#1}{\e_{#1}n_{#1}}{\e n}
}
\NewDocumentCommand{\nlk}{e_}{
\IfValueTF{#1}{n_{#1}l_{#1}k_{#1}}{nlk}
}
\NewDocumentCommand{\enlk}{e_}{
\IfValueTF{#1}{\e_{#1}\nlk_{#1}}{\e\nlk}
}
\NewDocumentCommand{\enl}{e_}{
\IfValueTF{#1}{\en_{#1}l_#1}{\en l}
}

\NewDocumentCommand{\nnl}{s}{
\IfBooleanTF{#1}{n_1 n_2 l}{n_1; n_2; l}
}
\NewDocumentCommand{\ennl}{s}{
\IfBooleanTF{#1}{\en_1 \en_2 l}{\en_1; \en_2; l}
}

\NewDocumentCommand{\gslm}{s}{
\IfBooleanTF{#1}{\sigma\lambda\mu}{\sigma;\lambda\mu}
}

\fi %

\ifdefined\COSMOMODELS
\else
\newcommand{\COSMOMODELS}{}
\usepackage{bm,upgreek}

\newcommand{\krn}[0]{\operatorname{k}}

\fi %

\newcommand{\DOS}{g}

\newcommand{\EDFT}{E}
\newcommand{\Eion}{E_{\text{ion}}}
\newcommand{\Eb}{E_{\text{b}}}
\newcommand{\Edc}{E_{\text{dc}}}
\newcommand{\SKS}{S}
\newcommand{\Tel}{T^{\mathrm{el}}}
\newcommand{\StrA}{{A}}
\newcommand{\StrB}{{M_j}}

\usepackage{graphicx}%
\usepackage{dcolumn}%
\usepackage{bm}%
\usepackage{hyperref}%
\usepackage[names]{xcolor}

\newcommand{\editor}[2]{%
  \expandafter\newcommand\csname #1note\endcsname[1]{%
    \textcolor{#2}{(\textbf{#1:} ##1)}}%
  \expandafter\newcommand\csname #1\endcsname[1]{%
    \textcolor{#2}{##1}}%
  \expandafter\newcommand\csname #1cancel\endcsname[1]{%
    \textcolor{#2}{\sout{##1}}}%
  \expandafter\newcommand\csname #1change\endcsname[2]{%
    \textcolor{#2}{\sout{##1} ##2}}%
  \newenvironment{#1text}{\color{#2}}{\color{black}}
}
\newcommand{\SM}{Supplemental Material}
\editor{MC}{blue}
\editor{FG}{teal}
\editor{CBM}{magenta}
\editor{resub}{cyan}

\begin{document}

\preprint{APS/123-QED}

\title{Predicting hot-electron free energies from ground-state data}%

\author{Chiheb Ben Mahmoud}%
\thanks{These authors contributed equally to this work}
\author{Federico Grasselli}
\thanks{These authors contributed equally to this work} 
\affiliation{%
Laboratory of Computational Science and Modeling, IMX, \'Ecole Polytechnique F\'ed\'erale de Lausanne, 1015 Lausanne, Switzerland
}%

\author{Michele Ceriotti}
\email{michele.ceriotti@epfl.ch}
\affiliation{%
 Laboratory of Computational Science and Modeling, IMX, \'Ecole Polytechnique F\'ed\'erale de Lausanne, 1015 Lausanne, Switzerland
}%

\date{\today}%

\begin{abstract}
Machine-learning potentials are usually trained on the ground-state, Born-Oppenheimer energy surface, which depends exclusively on the atomic positions and not on the simulation temperature.  
This disregards the effect of thermally-excited electrons, that is important in metals, and essential to the description of warm dense matter. An accurate physical description of these effects requires that the nuclei move on a temperature-dependent electronic free energy. 
We propose a method to obtain machine-learning predictions of this free energy at an arbitrary electron temperature using exclusively training data from ground-state calculations, avoiding the need to train temperature-dependent potentials, and benchmark it on metallic liquid hydrogen at the conditions of the core of gas giants and brown dwarfs.
This work demonstrates the advantages of hybrid schemes that use physical consideration to combine machine-learning predictions, providing a blueprint for the development of similar approaches that extend the reach of atomistic modelling by removing the barrier between physics and data-driven methodologies.

\end{abstract}

\pacs{Valid PACS appear here}%
\maketitle

In the past decade, machine learning (ML) algorithms proved to be an efficient alternative to expensive first principle (FP) calculations. The construction of ML interatomic potentials (MLIPs) trained on FP data has achieved a successful balance between computational cost and accuracy~\cite{Bartk2018, PhysRevLett.120.143001,behl11jcp,Kovcs2021}. This greatly simplified sampling the finite-temperature properties of materials, and has been complemented by ML models that predict functional materials properties ranging from scalar quantities~\cite{Paruzzo2018,Wang2021,Xie2018,pegolo2022temperature} to tensorial properties and fields~\cite{Grisafi2018,veit-dipoles,Chandrasekaran2019, CuevasZuvira2021}. %
Current ML strategies are usually designed to reproduce the ground state, Born-Oppenheimer (BO) potential energy surface, and do not account for the temperature-dependent electronic excitations which may play a major role in metallic matter at planetary conditions, like warm dense matter (WDM)~\cite{Scipioni2017,PhysRevB.104.144104,PhysRevLett.120.076401,Bonitz2020,McMahon2012}, and that introduce subtle but important corrections in the thermophysical properties of ordinary metals~\cite{grab+09prb,ma+15am}.
The most common strategy to treat finite electron temperature is to replace the BO potential with a temperature-dependent electronic free energy $A(\Tel)$.

In traditional MLIP frameworks, that rely exclusively on nuclear coordinates as inputs, switching from the BO potential to $A(\Tel)$ would require training a separate model for every target electronic temperature $\Tel$, recomputing also the training set -- although the temperature can be included as an input of the model, which yields MLIPs that are explicitly temperature-dependent, and interpolate between training data at different electron temperature~\cite{Zhang2020}.
One recent attempt to incorporate directly electronic excitations into ML simulations is to predict the single-particle density of states \cite{PhysRevB.102.235130}, and use it to evaluate a-posteriori corrections to the thermodynamic quantities, e.g. heat capacity or melting temperature, extracted from the MD of ions whose MLIP is trained on ground-state data~\cite{PhysRevMaterials.5.043802}. 
This approach is limited to condensed matter well below the Fermi temperature, where atomic forces are almost unaffected by the electronic excitations. 
Another recent method relies on Hamiltonian models based on the local density of states, and trained on finite temperature data~\cite{PhysRevB.104.035120}. Despite its success in describing directly electron finite temperature effects, this approach would still require generating data at specific target temperatures, limiting its transferability to conditions which span broad temperature ranges.%

In this  {Letter} we first show that, within a  {density functional theory (DFT)} framework, the total free energy, atomic forces and the stress tensor of the system can be rigorously approximated as the sum of a $\Tel=0$K contribution and a finite-$\Tel$ correction depending exclusively on the \emph{ground-state} electronic density of states (DOS).
This general result underpins a framework  that relies only on ground-state calculations to learn $A(\Tel)$ and its derivatives in the presence of thermally-excited electrons. Thus, a consistent ground-state training set and model can be generated, and used to sample the finite-electron-temperature distributions, using $\Tel$ as an external parameter. 
We test our method on simple metals, where the atomic forces are evaluated at increasing electronic temperatures.   
We then validate our framework by constructing the equation of state (EOS) of hydrogen at conditions relevant in gas giants and brown dwarfs, for temperatures up to 50,000K and pressures up to 1,600GPa, by means of MD simulations driven by our $\Tel$-dependent MLIPs. We also compute the heat capacity of hydrogen at 400GPa in the high temperature regime.

Let us start by considering the standard representation of the DFT energy:
\begin{equation}
\EDFT = \Eb - \Edc + \Eion
\end{equation}
as a sum of the electrostatic interactions between the ions $\Eion$, the band energy $\Eb = \sum_i f_i \epsilon_i$, expressed in terms of the Kohn-Sham (KS) eigenvalues $\epsilon_i$ and level occupations $f_i$, and the ``double-counting term"
\begin{equation}
\Edc = {\frac{1}{2} \iint \frac{\rho(\mathbf{r}^\prime)\rho(\mathbf{r})}{|\mathbf{r}-\mathbf{r}^\prime|} d\mathbf{r}d\mathbf{r}^\prime} - E_\mathrm{xc}[\rho]  + \int V_\mathrm{xc}[\rho](\mathbf{r}) \rho(\mathbf{r})d\mathbf{r}.
\end{equation}
Here $E_\mathrm{xc}$ is the is the exchange-correlation (XC) functional, $V_\mathrm{xc}[\rho](\mathbf{r}) = {\delta E_\mathrm{xc}}/{\delta \rho(\mathbf{r})}$ is the XC potential, and $\rho(\mathbf{r}) = \sum_i f_i |\phi_i(\mathbf{r})|^2$ is the DFT density, expressed in terms of the KS eigenfunctions $\phi_i(\mathbf{r})$ and occupations $f_i$.

\begin{figure}
    \centering
    \includegraphics{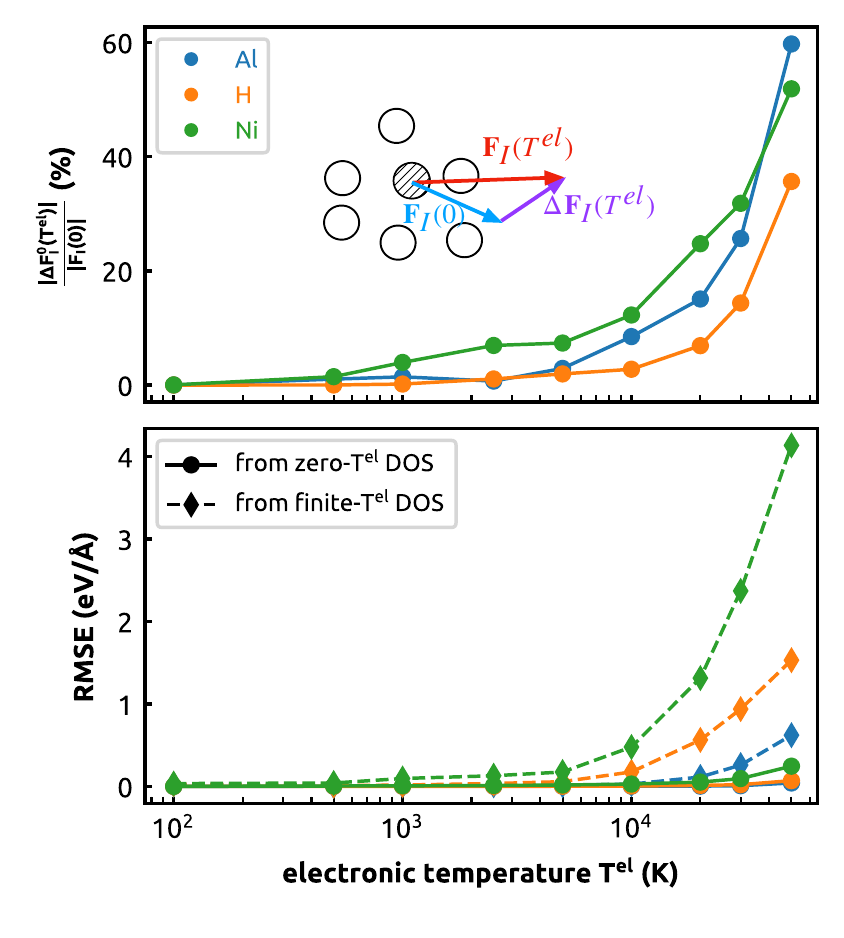}
    \caption{(Upper panel) Relative deviation of Hellmann-Feynman atomic force versus the electronic temperature with respect to the ground state force for a given ion and a Cartesian direction. The sketch represents the decomposition of the finite-$\Tel$ atomic force component within our framework. (Lower panel) Root mean square errors (RMSE) of 10 force components computed with Eq.~\eqref{eq:f-tel-terms} compared to their Hellmann-Feynman counterparts. Solid lines: using the DOS from $\Tel=0$K calculations. Dashed lines: using the DOS from finite-$\Tel$ calculations.
    Blue: aluminum; orange: hydrogen; green: nickel.
    }
    \label{fig:approx}
\end{figure}

Whenever an electronic temperature $\Tel$ is introduced, $f_i$ become fractional, and the correct energy functional becomes the Helmholtz free energy\cite{merm65pra,alav+94prl,marz-vand97prb}
\begin{equation}
    A(\Tel) = \EDFT(0) + \Delta \EDFT(\Tel) - \Tel\SKS(\Tel), \label{eq: Etot-TS}
\end{equation}
where $\Delta \EDFT(\Tel)$ is the finite-$\Tel$ contribution to the energy, and $S(\Tel)$
is the KS electronic entropy.  {From Eq.~\eqref{eq: Etot-TS} one can obtain the finite-$\Tel$ Hellmann-Feynman forces~\cite{Wentzcovitch1992}, whose relative deviation with respect to $\Tel=0$K-forces becomes significant at large $\Tel$, as reported in the upper panel of Fig.~\ref{fig:approx}.}
In principle, a $\Tel$-dependent XC functional should be employed~\cite{Karasiev2014}. However, it is often possible to rely on the Zero Temperature Approximation (ZTA), where the XC functional depends on $\Tel$ only through the $\Tel$-dependence of the density: $E_\mathrm{xc}[\rho(\Tel)]$. The ZTA performs well at both low and high $\Tel$ and also satisfies exact conditions as discussed in Ref.~\onlinecite{PribramJones2014}, and we adopt it as the basis of our framework.
 
A change in the occupation of the levels, e.g. as a consequence of thermal excitations, determines a change in the density, and thus, self-consistently, in the KS eigenenergies and eigenfunctions.
For instance, the  functional derivative of $\Eb$  with respect to $f_i$ is
\begin{equation}
\frac{\delta \Eb}{\delta f_i} = \epsilon_i + \sum_j f_j \frac{\delta \epsilon_j}{\delta f_i} \label{eq: variation Eband}
\end{equation}
Nonetheless, it can be proved, following a reasoning similar to that used in  {Ref.~\onlinecite{Weinert1992}, and in} Ref.~\onlinecite{Goedecker1992} for the energy variation due to infinitesimal atomic displacements (c.f. \SM~\cite{suppmat}), that the second term in Eq.~\eqref{eq: variation Eband} \textit{cancels exactly} with the variation of the double-counting term, ${\delta \Edc}/{\delta f_i}$.
Therefore, the change in $\EDFT$ due to a finite change in the occupations can be approximated by
\begin{equation}
\Delta \EDFT \approx \Delta \Eb^0 \equiv \sum_i \epsilon^0_i \, \Delta f_i ,
\label{eq: DeltaEtot}
\end{equation}
where $\epsilon^0_i \equiv \epsilon_i(\{\Delta f_k = 0\})$ are the eigenenergies computed at vanishing variation on all the $f_k$. The ``0'' superscript labels quantities  {obtained} from unperturbed eigenenergies, computed at $\Tel=0$K.

We now focus on the specific case where the set of $f_i$ are Fermi-Dirac distributed, $f_i = f(\tfrac{\epsilon_i-\mu(\Tel)}{k_B \Tel} )$,
$f(x)=1/(1+e^x)$ being the Fermi function, $\mu(\Tel)$ being the chemical potential of the electron system, and $k_B$ the Boltzmann constant.
From Eq.~\eqref{eq: DeltaEtot}, the finite-$\Tel$ correction to the DFT energy is
\begin{equation}
\Delta \Eb^0 (\Tel) = \int_{-\infty}^{+\infty} \epsilon \DOS^0(\epsilon) \left[
f\left(\tfrac{\epsilon-\mu(\Tel)}{k_B \Tel}\right) - f\left(\tfrac{\epsilon-\mu(0)}{k_B 0^+}\right)
\right] d\epsilon, \label{eq: Delta E integral}
\end{equation}
where $\DOS^0(\epsilon)=\sum_i \delta(\epsilon-\epsilon^0_i)$ is the electronic DOS.
 {$\mu(\Tel)$ is computed by enforcing charge-conservation:
\begin{equation}
    N = \int_{-\infty}^{+\infty} \DOS^0(\epsilon)\;
f\left(\tfrac{\epsilon-\mu(\Tel)}{k_B \Tel}\right) d\epsilon. \label{eq: equation for mu}
\end{equation}
In the \SM~\cite{suppmat}, we justify the use of $\DOS^0(\epsilon)$ in Eq.~\eqref{eq: equation for mu} and} 
when evaluating the electronic entropy $\SKS(\Tel)$
\begin{equation}
    \SKS(\Tel) \approx \SKS^0(\Tel) \equiv \int_{-\infty}^{+\infty} \DOS^0(\epsilon) \; s\left(\tfrac{\epsilon-\mu(\Tel)}{k_B \Tel}\right) d\epsilon, \label{eq: SKS continuum}
\end{equation}
where $N$ is the number of electrons and $s(x)= f \ln f + (1-f)\ln(1-f)$. 
Therefore, our approximation for the free energy reads: 
\begin{equation}
    A(\Tel) \approx E(0) + \Delta \Eb^0(\Tel) - \Tel \SKS^0(\Tel).
    \label{eq: A_approx}
\end{equation}
The number of states above $\mu(\Tel)$ that must be included to reliably compute the finite temperature contribution to the free energy depends on the temperature. This means that for training configurations we have to include a larger number of empty states than  {that} usually needed for $\Tel=0$K calculations. The finite-$\Tel$ correction terms in Eq.~\eqref{eq: A_approx} are independent of the alignment of the DOS (c.f. \SM~\cite{suppmat}), as long as it is chosen consistently when computing the chemical potential, the band energy and the entropy terms.
Our derivation justifies other approximations made in the literature such as the fixed-DOS approximation of Refs.~\onlinecite{Zhang2017}, which assumes that the electronic DOS is approximately independent of $\Tel$. In fact, the cancellations ensure the validity of Eq.~\eqref{eq: DeltaEtot}, even if the self-consistent energy levels (and thus the DOS itself) changed substantially by changing $\Tel$.
If one wanted to go beyond this ground-state approximation, it would not be sufficient to obtain the finite-$\Tel$ DOS, and to use it in expressions similar to Eqs.~\eqref{eq: Delta E integral}, \eqref{eq: equation for mu} and \eqref{eq: SKS continuum}. 
Without access to the self-consistent finite-temperature $\Edc$, doing so would lead to \emph{worse} results, as shown in the lower panel in Fig.~\ref{fig:approx}.
If one was prepared to perform self-consistent calculations at multiple temperatures, our perturbative expressions could also be applied to a reference temperature different from $\Tel=0$, and serve as the basis of more accurate temperature-interpolation schemes (c.f. \SM~\cite{suppmat}).
Our derivation directly translates to the calculation of derivatives of the free energy, like forces and stresses. For instance, according to the Born-Oppenheimer approximation, the force acting on the $I-$th nucleus in the the DFT ensemble is $\mathbf{F}_I(\Tel) = -\bm{\nabla}_I A(\Tel) \approx \mathbf{F}_I(0) + \Delta \mathbf{F}_I^0(\Tel)$ where
\begin{equation}
\begin{split}
    \mathbf{F}_I(0) \equiv & -\bm{\nabla}_I \EDFT(0)\\ \label{eq:f-tel-terms}
    \Delta \mathbf{F}_I^0(\Tel) \equiv & -\bm{\nabla}_I [\Delta \Eb^0(\Tel) - \Tel \SKS^0(\Tel)]. 
\end{split}
\end{equation}
In this decomposition, the electronic temperature $\Tel$ enters as an \textit{external parameter}.

These equations would be of limited practical value if the end goal was to compute $A(\Tel)$ for a given structure and temperature by means of a self-consistent electronic structure calculation. However, they become very useful in the context of data-driven modeling, as they provide a rigorous basis for the development of an ML framework to learn finite-$\Tel$ interatomic forcefields without the need to train on finite-$\Tel$ calculations.
The $\Tel=0$K quantities, $\EDFT(0)$ and $\mathbf{F}_I(0)$, can be modeled by any of the widely used MLIPs\cite{bartok2010,behler2011,PhysRevLett.120.143001,Schtt2018,Kovcs2021}. The hot-electron correction, Eq.~\eqref{eq:f-tel-terms}, can be accessed by training an ML model for the DOS. 
In this work we use the Gaussian Approximation Potentials (GAP)~\cite{bartok2010} and an atom-centered model for the DOS as detailed in Ref.~\onlinecite{PhysRevB.102.235130}. Both models rely on a linear expansion of a target quantity $y$ (be it the energy or the DOS) of a given structure $\StrA$ on a set of positive-definite functions, also called kernels, $\krn(\StrA,\StrB)$, measuring the similarity between the structure $\StrA$ and a structure $\StrB$ belonging to the set of $M$ reference environments, also called the active set:
\begin{equation}
    y(\StrA) = \sum_{j\in M} w_j \, \krn(\StrA,\StrB).
\end{equation}
Notice that, for the DOS, $y=g^0(\epsilon, \StrA)$, the weights $w_j(\epsilon)$ are a function of a discretized energy grid. The weights do not depend on the atomic positions of $\StrA$. Therefore, gradients needed to compute atomic forces, as in  Eq.~\eqref{eq:f-tel-terms}, do not act on the weights, but on kernels alone. In practice, since neither the kernels nor their gradients depend on $\epsilon$, they can be collected out of integrals involving the DOS like Eq.~\eqref{eq: Delta E integral}, which makes the implementation of this approach particularly simple in the case of kernel (or linear) models (c.f. \SM~\cite{suppmat}).
We construct the kernels $\krn(\StrA,\StrB)$ from the Smooth Overlap of Atomic Positions (SOAP) representation~\cite{bartok2013,Sandip} with radial scaling~\cite{will+18pccp} as implemented in \emph{librascal}~\cite{Musil2021}, and we use its  interface with i-PI~\cite{kapi+19cpc} to run MD simulations with finite electron temperature. 

\begin{table}
    \centering
        \begin{tabular}{l|c}
        \hline \hline
             & RMSE\\
             \hline
             $E(0)$&  $11.05$meV/atom \\ 
             $\Delta \Eb^0(\Tel) - \Tel \SKS^0(\Tel)$ &  $13.43$meV/atom\\
             $A(\Tel)$& $12.22$meV/atom \\
             $\mathbf{F}_I(0)$& $0.87$eV/\AA \\
             $\Delta \mathbf{F}_I(\Tel)$&  $0.66$eV/\AA \\
             $\mathbf{F}_I(\Tel)$&  $0.81$eV/\AA \\
             \hline \hline
        \end{tabular}
        \caption{Table of the validation root mean square errors (RMSE) of the ML models on the energies and forces compared to the reference  {DFT} data, at the same level of theory introduced in Eqs.~\eqref{eq: A_approx} and \eqref{eq:f-tel-terms}. The electronic temperature is $\Tel=35,000$K. The training set consists of 28,000 structures and the errors are reported for a validation set of 2,500 configurations.%
        }
        \label{tab: rmse_ml}
\end{table}

We demonstrate the practicality of our theoretical framework in ML workflows incorporating the electronic finite temperature effects in atomistic simulations by constructing the EOS of metallic liquid hydrogen %
at conditions similar to those found in the core of a young Jupiter \cite{Nettelmann2012}, and we compare our ML approach to explicit first-principles molecular dynamics (FPMD) simulations results at finite-$\Tel$. 
We build a training set made of $\sim 28,000$ structures, each containing $128$ atoms, and densities ranging between 0.6g cm$^{-3}$ and 1.77g cm$^{-3}$. It consists of configurations from Ref.~\cite{Cheng2020}, complemented by snapshots obtained from  MD simulations performed with preliminary versions of the MLIP. We employ \textsc{Quantum ESPRESSO}~\cite{qe2009,Giannozzi2017,Giannozzi2020} (QE) for DFT calculations of the data set, using the Optimized Norm-Conserving Vanderbilt pseudopotential~\cite{Schlipf2015} version 1.2, which is shown to perform well even at $\sim$TPa pressures~\cite{sun+15ncomm}. 
Dispersion interactions are included via a van der Waals density functional~\cite{Berland2015,Thonhauser2015,Langreth2009,Thonhauser2007}. 
Details of the reference calculations and of the GAP model construction are given in the \SM~\cite{suppmat}. 
Table.~\ref{tab: rmse_ml} shows the root mean square error (RMSE) of the different (free) energies and forces for $\Tel=0$K and at $\Tel=35$,$000$K. The ML models are in good agreement with the corresponding DFT calculations and the RMSE of the total free energy is well below the typical thermal energy at the temperatures we consider in this study, 
and comparable to the values observed in previous simulations of liquid systems at high ionic temperatures~\cite{deri-csan17prb}. 

\begin{figure}
    \centering
    \includegraphics{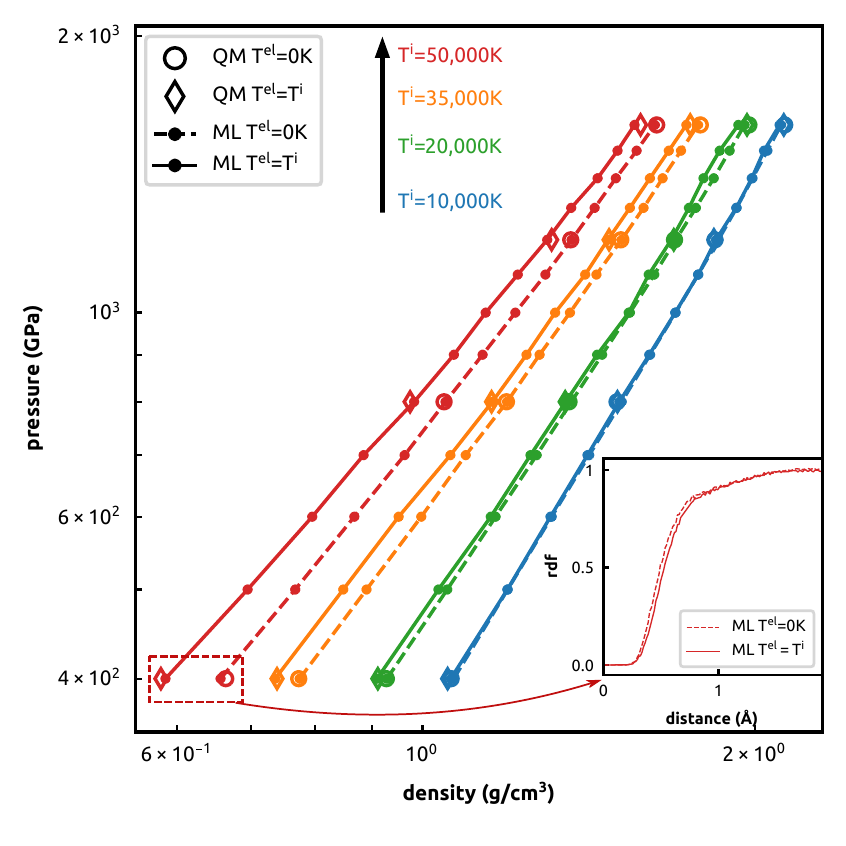}
    \caption{Hydrogen isotherms of different equations of state (EOS). The empty circles correspond to the EOS computed with ``cold''-electron-$\Gamma$-point DFT. The empty diamond correspond to the EOS computed with finite-electron-temperature-$\Gamma$-point DFT. The  {dashed} lines correspond to the EOS computed with ML trained on $\Tel=0$K data. The  {solid} lines correspond to the EOS computed with finite temperature ML framework, where $\Tel=T^i$. The temperatures  {range} from 10,000K to 50,000K denoted by the different colors as shown in the  {legend}. The statistical error bars computed by block averages are smaller than the size of the markers. The small inset represents the radial distribution functions of hydrogen at $P=400$GPa and $T^i=$ 50,000K computed from the ML trajectories. The  {dashed} line corresponds to $\Tel=0$K and the  {solid} line to $\Tel=T^i$.
}
    \label{fig:EOS}
\end{figure}

In order to gauge the importance of finite-$\Tel$ effects, and to obtain accurate reference calculations consistent with our computational setup, we run two sets of FPMD trajectories targeting the pressures 400GPa, 800GPa, 1,200GPa, and 1,600GPa for each of the ionic temperatures $T^i=$ 10,000K, 20,000K, 35,000K, and 50,000K. The electronic temperature of the first set is $\Tel=0$K, while $\Tel=T^i$ in the second set. The DFT calculations are performed with QE and $\Gamma$-point sampling. We evolve the ion dynamics with i-PI for at least 8ps, after an equilibration of 1ps, with a time step of 0.1fs. $T^i$ is controlled by stochastic velocity rescaling\cite{buss+07jcp} with a time constant $\tau=5$fs, and an isotropic barostat\cite{buss+09jcp} with a time constant $\tau=20$fs, thermalized with an optimal-sampling generalized Langevin thermostat\cite{ceri+10jctc}.
Due to the high temperature and the fast intrinsic time scale of hydrogen, such relatively short simulations are sufficient to obtain converged results with a small statistical uncertainty.    
We report the results of these simulations in Fig.~\ref{fig:EOS}, by the empty  {symbols}. The differences due to the finite electron temperature grow steadily between 10,000K and 50,000K, and at the highest temperature they range between  $4\%$ at 1,600GPa and $10\%$ at 400GPa, providing an indication of the impact of finite-$\Tel$ in this range of pressure and density. %

\begin{figure}
    \centering
    \includegraphics{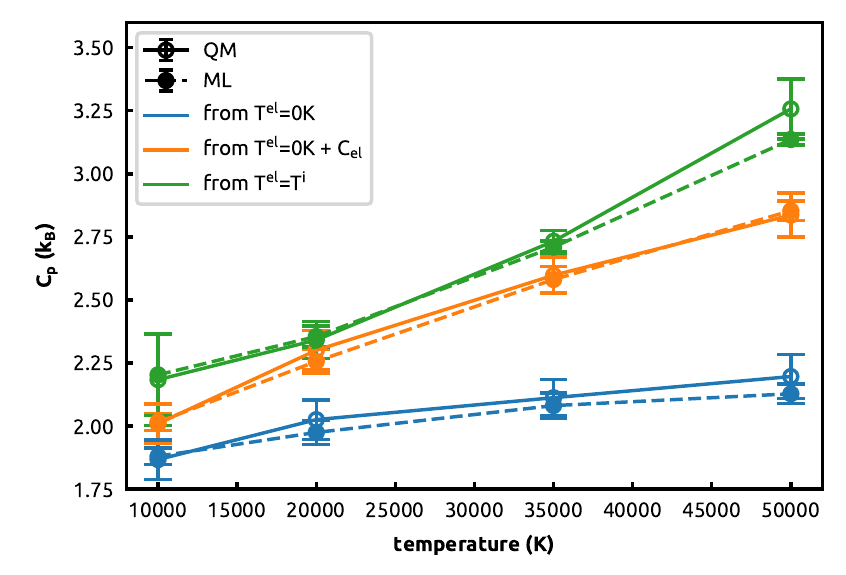}
    \caption{Specific heat capacity $C_p$ of hydrogen from $NpT$ simulations at $400$GPa. The solid lines represent the  {DFT} calculations and the dashed lines represent the ML calculations. Blue: $C_p$ from the fluctuations of the ions' enthalpy at $\Tel=0$K; orange: $C_p$ same as the blue curves in addition to a correction term computed from the average band energy of the electrons over the trajectories; green: $C_p$ from the the finite-$\Tel$ sampling. The error bars are computed from standard block analysis.}
    \label{fig:Cp}
\end{figure}

We then run two analogous sets of trajectories based on the finite-$\Tel$ MLIP, temperatures as for the FPMD, and pressures  spanning the range between 400GPa and 1,600GPa in intervals of $100$GPa. As for the case of FPMD, the first set of simulations does not include any finite temperature effects ({dashed} lines in Fig.~\ref{fig:EOS}), while the second incorporates them ({solid} lines). %
Our ML EOSs are in excellent agreement with the reference curves obtained with explicit finite-$\Tel$ FPMD, up to the statistical uncertainties. We also observe a small shift in the radial distribution at the lower pressure and higher temperature range,  corresponding to the difference in  particle densities.
As an additional demonstration of the importance of incorporating finite-$\Tel$ effects, we compute constant-pressure heat capacities, $C_p = \left(\frac{\partial H}{\partial T}\right)_p$ that we obtain as finite differences of the enthalpy 
$H= \left< K \right> + \left<A(\Tel)\right>+\Tel \left<S^0(\Tel)\right> + p\left< V\right> $, as described in the \SM~\cite{suppmat}. 
Here $ K $ is the kinetic energy of the ions, and the averages $\left< \ldots \right>$ are computed over finite-$\Tel$ $NpT$ sampling.
Fig.~\ref{fig:Cp} compares the heat capacity computed from $\Tel=0$K simulations (blue) with that computed including the electronic contributions (green) - which amounts to almost 50\%{} at the highest temperature considered.  {DFT} and ML simulations agree with each other within their statistical uncertainty. 
The a-posteriori incorporation of electronic excitation by adding $C_{el} \equiv \left< \frac{\partial \Delta\Eb}{\partial T} \right>$  (orange) on top of the $\Tel=0$K ionic contribution, as done in Ref.~\onlinecite{PhysRevMaterials.5.043802}, cannot reproduce accurately the finite-$\Tel$ results. 

These results demonstrate the accuracy of an ML model based on the ground-state DOS approximation in sampling the finite-$\Tel$ thermophysical properties of hydrogen in a challenging portion of its phase diagram.
By treating explicitly the ionic and electronic degrees of freedom, our ML models eliminate one of the most glaring limitations of traditional MLIPs, that are restricted to perform simulations at a single (usually zero) electron temperature. 
We remark that no restriction occurs in applying our machinery to a two-temperature model where $\Tel \neq T^i$ and the hot electrons are not in thermal equilibrium with the nuclei, even though electron-nuclei interactions should be included to allow for thermalization.
Our approach can be easily extended to any electronic structure method based on the KS mapping, and can be naturally used also for multiple-species systems, opening the possibility of studying the complex phase diagram of metallic mixtures at high-pT conditions, which dictates the evolution of giant planets \cite{Helled2020}. 
On a conceptual level, the idea of using a physical approximation in synergy with data-driven predictions indicates a promising research direction to further extend the scope of applicability of predictive atomic-scale simulations.  
\begin{acknowledgments}
We thank G. Mazzola for discussing with us an early version of the manuscript, and G. Gil for useful comments. CBM and MC acknowledge support by the Swiss National Science Foundation (Project No. 200021-182057) and the NCCR MARVEL, a National Centre of Competence in Research, funded by the Swiss National Science Foundation (grant number 182892).
FG acknowledges funding from the European Union's Horizon 2020 research and innovation programme under the Marie Sk\l{}odowska-Curie Action IF-EF-ST, grant agreement no. 101018557 (TRANQUIL).
\end{acknowledgments}

 {\emph{Data availability}: The data supporting the findings in this work are available on the Materials Cloud platform\cite{talirz2020materials} at DOI:10.24435/materialscloud:36-ff. Archived versions of software used to run the DFT and ML simulations are available on Zenodo\cite{zenodo} at DOI:10.5281/zenodo.7044489.}
\end{document}